# Hard antinodal gap revealed by quantum oscillations in the pseudogap regime of underdoped high-$T_\mathrm{c}$ superconductors

Máté Hartstein,[1] Yu-Te Hsu,[1][*] Kimberly A. Modic,[2][†] Juan Porras,[3]
Toshinao Loew,[3] Matthieu Le Tacon,[3,4] Huakun Zuo,[5] Jinhua Wang,[5]
Zengwei Zhu,[5] Mun K. Chan,[2] Ross D. McDonald,[2] Gilbert G. Lonzarich,[1]
Bernhard Keimer,[3] Suchitra E. Sebastian,[1] Neil Harrison[2]

[1]Cavendish Laboratory, Cambridge University, J.J. Thomson Avenue, Cambridge CB3 OHE, UK

[2]Los Alamos National Laboratory, Los Alamos, Mail Stop E536, Los Alamos, NM 87545, USA

[3]Max Planck Institute for Solid State Research, Heisenbergstr. 1,D-70569 Stuttgart, Germany

[4]Karlsruhe Institue of Technology, Institute for Quantum Materials and Technologies,
Hermann-v.-Helmholtz-Platz 1, D-76344 Eggenstein-Leopoldshafen, Germany

[5]National High Magnetic Field Center and School of Physics, Huazhong University of
Science and Technology, Wuhan 430074, China

[*]current address: High Field Magnet Laboratory (HFML-EMFL) and Institute for Molecules and Materials, Radboud University, Toernooiveld 7, 6525 ED Nijmegen, Netherlands

[†]current address: Institute of Science and Technology Austria, Am Campus 1, 3400 Klosterneuburg, Austria

**An understanding of the missing antinodal electronic excitations in the pseudogap state is essential for uncovering the physics of the underdoped cuprate high temperature superconductors. The majority of high temperature experiments performed thus far, however, have been unable to discern whether the antinodal states are rendered unobservable due to their damping, or whether they vanish due to their gapping. Here we distinguish between these two scenarios by using quantum oscillations to examine whether the small Fermi surface pocket, found to occupy only 2% of the Brillouin zone in the underdoped cuprates, exists in isolation against a majority of completely gapped density of states spanning the antinodes, or whether it is thermodynamically coupled to a background of ungapped antinodal states. We find that quantum oscillations associated with the small Fermi surface pocket exhibit a signature sawtooth waveform characteristic of an isolated two-dimensional Fermi surface pocket. This finding reveals that the antinodal states are destroyed by a hard gap that extends over the majority of the Brillouin zone, placing strong constraints on a drastic underlying origin of quasiparticle disappearance over almost the entire Brillouin zone in the pseudogap regime.[7–18]**

We observe key experimental evidence in the underdoped high-$T_c$ superconductor $YBa_2Cu_3O_{6.55}$ for a Fermi surface comprising a small isolated two-dimensional pocket in which the majority of electronic density of states is fully gapped, by our observation of a de Haas-van Alphen (dHvA) oscillation waveform with striking similarities to that exhibited by ideal two-dimensional metals.[25–32] The shape of the quantum oscillation waveform enables a key distinction to be made between a Fermi surface comprising multiple sections, and one comprising an isolated Fermi surface section. We can thus identify whether the Fermi surface of the underdoped cuprates corresponds to a large paramagnetic Fermi surface reconstructed into multiple Fermi surface sections (Fig. 1a,b),[33–36] or a small isolated Fermi surface section, the remainder having been gapped (Fig. 1c,d). This distinction can be made on the basis of (i) the direction of the sawtooth quantum oscillation waveform in magnetisation (and magnetic torque), whether forward-leaning in direction, or reverse-leaning in direction (i.e. an inverse sawtooth); (ii) the shape of the quantum oscillation waveform in the magnetic susceptibility (and the electrical resistivity), whether an 'inverted U' shape, or a 'U' shape; and (iii) the relative amplitude and sign of each successive harmonics.[28,37,38] These signatures can only be identified if adequate harmonic content is observed in the measured quantum oscillations, which is a direct consequence of sample purity combined with signal detection threshold.

Figure 2a (see also Supplementary Figs. 1 and 2) shows the observation of five Fourier harmonics in the present single crystals of underdoped $YBa_2Cu_3O_{6+x}$, revealed to be of substantially higher quality than the previous generation of samples in which only three Fourier harmonics are observed.[39] The large harmonic content (see Fig. 2a) in our present samples is a consequence of low impurity (Dingle) damping, which reflects a narrow Landau level width from low defect and impurity scattering. Fig. 2b shows the substantially larger quantum oscillation amplitude in the present measurements (green) compared to measurements on previous samples[23,39] (here the comparison is made by rescaling amplitudes to match in the limit $F/B \to 0$ using the extrapolation made in Fig. 2a). To discern whether the waveform characteristics correspond to an isolated two-dimensional Fermi surface or a Fermi surface comprising multiple sections, we focus on the region above $\approx 48$ T where the largest number of harmonics are observed (Fig. 3a,b). An exponential fit to the magnetic field dependence of the peak-to-peak quantum oscillations from Fig. 2d (shown by the solid line in Fig. 2c) reveals a considerably smaller effective Dingle damping of $\Gamma' \approx 83$ T (see Supplementary Information) for the present samples, compared to the higher Dingle damping of $\approx 140$ T previously obtained for samples measured in refs.[23,39] When the applied field is orientated along the crystalline $c$-axis the single

Fermi surface pocket gives rise to multiple frequency components due to interlayer coupling and magnetic breakdown effects.[39] For the purposes of simple waveform comparison, the majority of measurements reported here are made by inclining the crystalline *c*-axis of the sample by ≈ 36° with respect to magnetic field direction (Fig. 2d). At this angle, angle-resolved quantum oscillation studies have shown the waveform to consist of only a single fundamental frequency for which the near degeneracy of its spin-up and -down Landau levels leads to a spin-damping factor of $R_s \approx 1$ (see Supplementary Information).[39] Fig. 3a shows the dHvA waveform in $YBa_2Cu_3O_{6.55}$, which is seen to have a forward-leaning sawtooth form - the direction of the sawtooth form is identified from the direction of the diamagnetic response in measured magnetic torque (Supplementary Fig. 3). The observed sawtooth form is seen to be similar to that measured in the well known ideal two-dimensional electron gas in a GaAs/AlGaAs heterostructure (Fig. 3c).[29] Fig. 3 shows all the hallmarks of a clean two-dimensional metal consisting of a single section of Fermi surface in $YBa_2Cu_3O_{6.55}$, including (i) a forward leaning 'sawtooth' waveform in the magnetisation and the magnetic torque (Figs. 3a and c), (ii) an inverted 'U'-shape waveform in the magnetic susceptibility and the resistivity (Figs. 3a and c), and (iii) multiple harmonics whose amplitudes fall on an approximately exponential curve (Figs. 3b and d). The forward-leaning sawtooth dHvA oscillations observed in $YBa_2Cu_3O_{6.55}$ are inconsistent with the inverse sawtooth expected for a large reservoir contribution ($\zeta_{res}$) to the electronic density of states at the Fermi level (Supplementary Information and Supplementary Figs. 4 and 5).

Our observation of forward-leaning sawtooth dHvA oscillations in underdoped $YBa_2Cu_3O_{6+x}$ points to an isolated two-dimensional Fermi surface section. The forward-leaning sawtooth shape of the dHvA quantum oscillatory waveform arises from the jumps in chemical potential between discrete filled and empty Landau levels as the magnetic field is swept.[25,27,28,30,32] In contrast, a Fermi surface section coexisting with a significant reservoir contribution ($\zeta_{res}$) to the electronic density of states at the Fermi level, for instance from other Fermi surface sections, would cause notable departures from the forward-leaning sawtooth waveform expected for ideal two-dimensional metals (Supplementary Information and Supplementary Fig. 4), ultimately resulting in a backward leaning sawtooth waveform for a sufficiently large reservoir contribution. A large reservoir contribution to the electronic density of states at the Fermi level would be expected for scenarios such as those involving additional Fermi sections,[40] open Fermi surface sheets,[28] a significant interlayer dispersion of the Fermi surface,[28,37,38,41] an additional broad incoherent density of states,[29] and others (see Supplementary Information).

In order to place an upper bound on the relative size of any reservoir accompanying the observed Fermi surface pocket, we make quantitative comparisons of the amplitude and sign of each successive harmonic with model predictions in which the size of the reservoir is treated as a variable (Fig. 4). We model the Fermi surface in $YBa_2Cu_3O_{6.55}$ by a single two-dimensional Fermi surface section accompanied by a finite reservoir of electronic density of states at the Fermi level. In the limit of an ideal two-dimensional metal characterised by an isolated Fermi surface section, each of the harmonics contributing to the waveform has an amplitude that decreases in an exponential fashion with increasing harmonic index $p$. For $T \to 0$, the susceptibility takes the form

$$\frac{\partial M}{\partial B} \propto -\sum_p a_p \cos\left(\frac{2\pi p F_0}{B}\right), \tag{1}$$

where $a_p = e^{-\frac{p\Gamma'}{B}} R_{T,p}$, $F_0$ is the fundamental quantum oscillation frequency, $R_{T,p}$ is a thermal damping factor close to unity at $T = 1.5$ K and we have assumed $R_s = 1$ (see Supplementary Information).[37] Here $\Gamma'$ is the effective Dingle damping term that parametrizes damping caused by quasiparticle scattering and other effects, and is related to the effective scattering rate $\tau^{-1}$ via $\tau^{-1} = e\Gamma'/\pi m^*$ (see Supplementary Information).[28] The measured quantum oscillation data (Figs. 2a,c and Fig. 3b) show an exponential decrease in the amplitude of successive harmonics

– all of which have the same sign, as expected for an ideal two-dimensional metal. In contrast, a finite reservoir contribution $\zeta_{\text{res}}$ to the electronic density of states would yield consecutive harmonic amplitudes that deviate from an exponential decrease, and exhibit relative changes in harmonic sign for a sufficiently large reservoir size, starting with the higher harmonics (Supplementary Information).[28] We can thus place an upper bound on the size of any finite reservoir contribution by comparing the relative amplitude and size of each successive measured harmonic with a model simulation in which the reservoir size is varied.

In Fig. 4, we simulate a Fermi surface model in which a two-dimensional Fermi surface with an energy-averaged or constant component of electronic density of states at the Fermi level $\zeta_{\text{2D},0}$ is accompanied by a finite reservoir with electronic density of states $\zeta_{\text{res}}$ (Supplementary Information). Fig. 4 shows a simulation of the relative magnitude and sign of consecutive harmonics for such a Fermi surface model in which the reservoir size is varied (solid curves). Horizontal dotted lines show the magnitude and sign of the five leading experimentally observed harmonics. We identify for which reservoir size the magnitude and sign of each consecutive harmonic best agrees with the experimental measurements, and find that the best agreement is for a value of the ratio $\zeta_{\text{res}}/\zeta_{\text{2D},0} \approx 0$ (indicated by squares in Fig. 4). This absence of a reservoir electronic density of states (i.e. $\zeta_{\text{res}} = 0$) is consistent with the exponential decrease of the successive harmonic amplitude size shown in Fig. 2. The rigour of our identification of an isolated Fermi surface in $YBa_2Cu_3O_{6+x}$ is further established by the agreement in size of *all* of the simulated harmonic amplitudes with those calculated for a value of the effective Dingle damping factor $\Gamma' = 83$ T (extracted from the magnetic field-dependence of peak-to-peak quantum oscillation amplitude in Fig. 2c). The upper bound of the reservoir density of states is set by the error bar in the comparison made in Fig. 4, which is found to be $\approx 0.03 \times \zeta_{\text{2D},0}$ (shaded region in Fig. 4). Our analysis of the dHvA effect is a particularly robust probe of the reservoir density of states at the Fermi level since the sign of each of the harmonics is independent of the value of the effective Dingle damping factor (different values shown in model simulations in the Supplementary Information), making our conclusion of a vanishingly small $\zeta_{res} \ll \zeta_{2D,0}$ unaffected by uncertainties in the value of the effective Dingle damping factor used for model simulations.

Complementary experimental observations are in good agreement with our finding of a Fermi surface consisting of an isolated electron pocket. The component of the linear coefficient of the electronic heat capacity associated with the Fermi surface at high magnetic fields has been measured to have a value $\gamma \approx 4 \pm 1$ mJ·mol$^{-1}$K$^{-2}$,[42,43] in high magnetic fields, which is close to the value $\gamma = 2 \times 1.47 \times m^* \approx 4.7$ mJ·mol$^{-1}$K$^{-2}$ expected for a single Fermi surface pocket per CuO$_2$ plane with effective mass $m^* \approx 1.6\ m_e$ (where $m_e$ is the free electron mass).[23] The value of the Hall coefficient at high magnetic fields[44] is also found to be consistent with a single electron pocket per CuO$_2$ plane.[45,46] While a maximum reservoir electronic contribution to the heat capacity of $\gamma \approx 0.1$ mJ·mol$^{-1}$K$^{-2}$ is set by the upper limit of the reservoir electronic density of states $\zeta_{res}$ that we find in Fig. 4, our complementary measurements rule out the possibility of a contribution from additional small light Fermi surfaces that have been suggested by models shown in schematic Figs. 1a and b.[47,48] Were additional light Fermi surface sections to be present, their light effective mass ($m^*_{hole} = 0.45\ m_e$)[46,47] and multiplicity in the Brillouin zone (seen in schematic Figs. 1a and b) would yield a dominant contribution to the dHvA signal (Supplementary Information), and would also cause a significant departure from a forward-sloping sawtooth waveform. Yet both signatures are excluded by our dHvA measurements (Supplementary Information and Supplementary Fig. 6), ruling out such additional small light Fermi surface sections accompanying the isolated pocket that yields the observed quantum oscillations.

Our finding of an isolated Fermi surface pocket occupying 2% of the Brillouin zone leads to the conclusion that in the pseudogap ground state, the majority of the density of states at the Fermi level have been eliminated by a hard gap. The location of this pocket is indicated by comparison with complementary momentum dependent probes, such as photoemission and interlayer conductivity,[1–6,10–12] which find the majority antinodal electronic density of states at the Fermi level to be absent, while coherent quasiparticle states are observed primarily in the nodal regions.[5,6,42] A scenario in which the antinodal states are unobserved due to damping is ruled out by our measurements, given the substantial residual density of states that would be expected in such a scenario. Our quantum oscillation measurements thus point to a complete gapping of the density of states at the Fermi level over the majority of the Brillouin zone, leaving behind an isolated nodal density of states yielding the observed Fermi surface section.[23,39] A single nodal Fermi surface section (per CuO plane) is sufficient to account for the finite electronic density of states identified in heat capacity and nuclear magnetic resonance experiments in strong magnetic fields.[42,43,49,50]

Starting with a large paramagnetic band structure Fermi surface, a leading possibility for a nodal Fermi surface is from Fermi surface reconstruction by a charge-density wave enhanced by the high magnetic fields observed by nuclear magnetic resonance and X-ray diffraction experiments.[7,33,35,36,51] In such scenarios, however, the strength of the order parameter appears to be too weak to gap the antinodal density of states, instead resulting in multiple sections of the Fermi surface, as shown in schematic Figs. 1a and b.[35,36] Possibilities for an antinodal gapping of density of states at the Fermi level include antiferromagnetism,[52] topological order (short range antiferromagnetism),[53] spin liquid phases,[12] valence bond order,[54] and staggered flux phases,[7,11] which are proposed to produce a hard antinodal gap for a sufficiently large potential. The electronic contribution to the density of states from the nodal Fermi surface pockets produced by these models would need to be reconciled with that seen in heat capacity experiments.[23] Hard antinodal gapping of the density of states at the Fermi level may also be yielded by a strong superconducting gap with a *d*-wave pairing manifold that persists under strong magnetic fields,[55] and participates in a composite order parameter to yield a nodal Fermi surface pocket in conjunction with a hard antinodal gap. For instance, a pair-density wave has been suggested to characterise the pseudogap,[56] in which $d$-wave superconductivity coexists with a charge-density wave,[57–60] yielding a nodal Fermi surface from Fermi surface reconstruction. Any pseudogap model must produce a hard gap that destroys the majority density of states at the Fermi level spanning the antinodal region in momentum space.

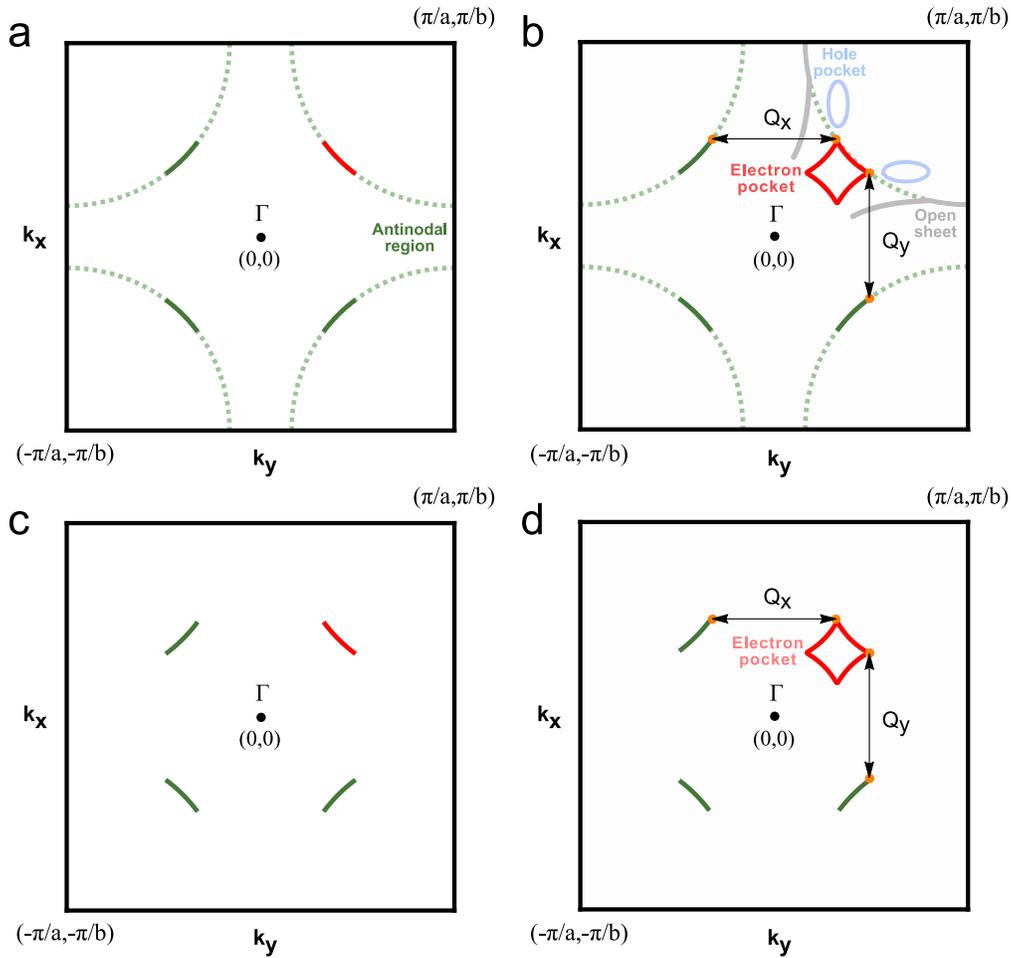

Figure 1: **Two proposed scenarios of the pseudogap ground state in $YBa_2Cu_3O_{6+x}$.** (**a**) Fermi arcs are part of a large cylindrical Fermi surface, while the antinodal states are rendered incoherent. (**b**) Fermi surface translated for example by biaxial charge density wave order vectors, $\mathbf{Q}_x$ and $\mathbf{Q}_y$, and reconstructed into a nodal electron pocket, antinodal hole pockets and one-dimensional open sheets. (**c**) Alternatively, Fermi arcs are unusual isolated objects sharply truncated in momentum space, while the remaining density of states encompassing the antinodal region is completely gapped. (**d**) Fermi arcs connected for example by biaxial charge density wave order vectors yield a single isolated nodal electron pocket. While Fermi surface reconstruction by a biaxial charge density wave has been chosen as an illustration,[45] alternative models of Fermi surface reconstruction could potentially yield similar results.

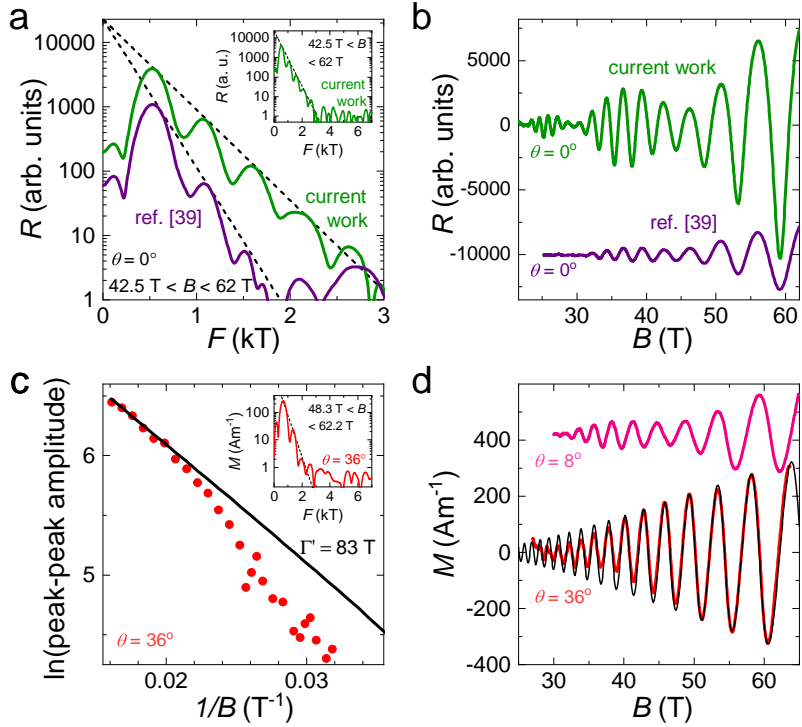

Figure 2: **Enhanced quantum oscillations compared to measurements on previous samples.** (**a**) The Fourier transform shows rich harmonic content of measured quantum oscillations (shown in **b**). Dashed lines show the smaller exponential amplitude damping with increasing harmonic index for current samples compared to those measured in ref.,[39] signalling a substantially lower impurity (Dingle) damping for the new generation of samples. The same Fourier transform is shown in the inset with an extended frequency range. (**b**) Comparison of absolute amplitude of quantum oscillations in contactless resistivity measured as a function of magnetic field reveals considerably larger quantum oscillations (green) measured on present samples compared to previous measurements (purple)[39] (data are scaled to have equal amplitude in the infinite magnetic field limit). Throughout, $B = \mu_0 H_0$, where $H_0$ is the applied magnetic field. (**c**) Peak-to-peak amplitude of the dHvA oscillations as a function of magnetic field (red circles) compared with exponential simulation $e^{-\frac{\Gamma'}{B}}$ (black line), from which an effective Dingle damping factor of $\Gamma' = 83$ T is estimated. The observed increase in damping at lower magnetic fields could arise from effects such as a small nodal gap (see Supplementary Information). The inset shows the Fourier transform of the dHvA oscillations. (**d**) Oscillations in the magnetisation (measured in the magnetic torque) at two different angles (magenta and red), showing a single series of oscillations when $\theta = 36°$. The black line shows simulated dHvA oscillations for an isolated two-dimensional Fermi surface (see Supplementary Information), assuming a quasiparticle effective mass $m^*$, where $m^* \cos\theta = 1.6\, m_e$ and $m_e$ is the free electron mass.

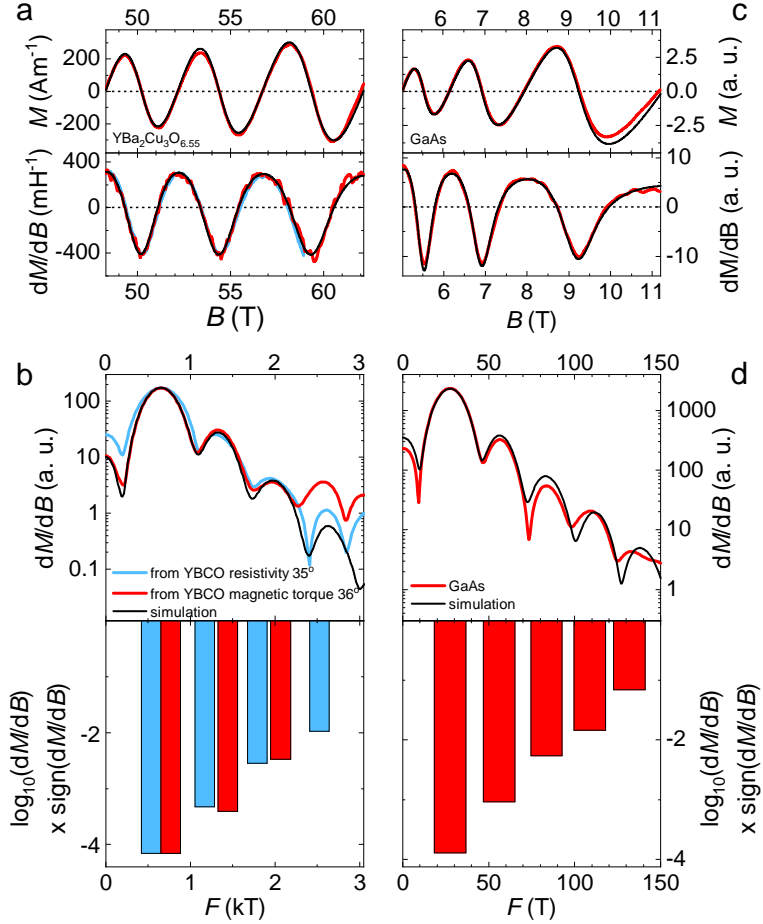

Figure 3: **Signature quantum oscillations from an isolated two-dimensional Fermi surface with no reservoir density of states at the Fermi level.** (**a**) de Haas-van Alphen oscillations in the magnetisation ($M$, upper panel, as inferred from magnetic torque, and susceptibility ($\partial M/\partial B$, lower panel) of YBa$_2$Cu$_3$O$_{6.55}$ for $B$ inclined at an angle of 36° from the crystalline $c$-axis (red curve) compared to numerical simulations in black. Light blue curves show resistivity data rescaled by $B^{-2}$ yielding a comparable quantity to the magnetisation.[25] Throughout, $B = \mu_0 H_0$, where $H_0$ is the applied magnetic field. (**b**) Corresponding Fourier transform amplitude (upper panel) and relative logarithmic Fourier transform amplitude of successive harmonics multiplied by the sign of each harmonic (lower panel). Black curves represent a numerical simulation (Supplementary Information), while red curves are extracted from experimental magnetic torque data, and light blue curves are extracted from experimental resistivity data. (**c**) Shape of the de Haas-van Alphen oscillation waveform and harmonic content of the magnetisation ($M$) and susceptibility ($\partial M/\partial B$) of a GaAs heterostructure[29] closely resembles that of YBa$_2$Cu$_3$O$_{6.55}$. (**d**) Corresponding Fourier transform amplitude of successive harmonics and relative logarithmic Fourier transform amplitude multiplied by the sign of each harmonic for GaAs. The forward-leaning sawtooth waveform and linear decrease of the logarithmic amplitude of successive harmonics in both materials is consistent with the expectation for an isolated two-dimensional Fermi surface.

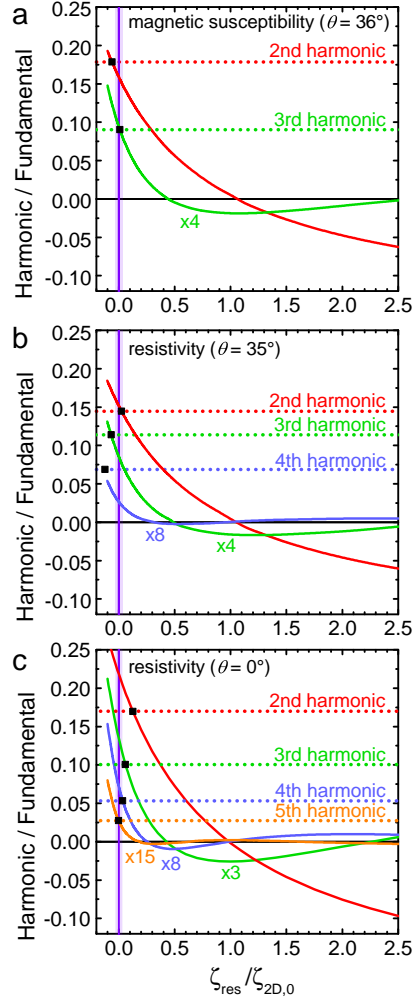

Figure 4: **Reservoir contribution $\zeta_{\text{res}}$ to the electronic density of states at the Fermi level from harmonic analysis.** Quantum oscillations in magnetic susceptibility and resistivity are simulated as a function of the size of the reservoir density of states using an effective Dingle damping factor of $\Gamma' = 83$ T, and decomposed into harmonic contributions. The simulated amplitudes of each of the harmonics relative to the fundamental are shown by solid curves. Black squares mark the experimental values. The experimentally measured magnitude and sign of consecutive harmonics (dashed horizontal lines) best agrees with model simulations for a complete absence of reservoir density of states, with an upper limit of 3% of the electron mass given by the error bar (purple shade).

## Methods

**Details of the sample growth.** Single crystals of $YBa_2Cu_3O_{6+x}$ were grown by the self-flux method.[61,62] Oxygen content of $x = 0.55$ was set by placing the crystals in a furnace with controlled oxygen partial pressure for a week. All crystals used in this study were mechanically detwinned and ortho-II ordered. Magnetisation characterisation at 2 Oe were used to determine the superconducting transition temperature $T_c$ of 61 K, taken as the mid-point value, and transition widths $\Delta T_c$ (10% - 90%) of within 1.5 K. The obtained $T_c$ confirms the oxygen content of $YBa_2Cu_3O_{6.55}$ and corresponds to a hole doping of 0.108.[61] Samples were screened at high magnetic fields to identify single crystals with the lowest damping $\Gamma\prime$ (Fig. 2).

**Details of dHvA measurements.** dHvA oscillations are measured in the magnetic torque $\tau = \mathbf{m} \times \mathbf{B}$ using the 65 T pulsed field magnet in the National High Magnetic Field Laboratory (NHMFL), Los Alamos. The magnetic torque was measured using an piezoelectric cantilever by inclining the crystalline $c$-axis of the sample at angle $\theta$ with respect to the magnetic field, where $\sin \theta = \mathbf{m} \cdot \mathbf{B}/|\mathbf{mB}|$. The sign of the magnetic torque is determined by measuring the deflection of the cantilever in response to the gravitational force of the sample (determined by rotation by $\theta = 180°$), and by measuring the sign of the magnetic hysteresis caused by pinning of vortices, which gives rise to a large deflection especially at low temperatures (see Supplementary Fig. 3). The component of the magnetisation moment along the magnetic field direction is given by $|M| \propto |\tau|/|\mathbf{B}|$. The first to third harmonics are observed as robust features in the Fourier transform, appearing at discernibly higher values than the noise for signal-averaged magnetic torque data.

**Details of quantum oscillations measured using the resonant oscillator technique.** Contactless resistivity was measured via the proximity detector oscillator (PDO) technique using the 65 T pulsed field magnet in the National High Magnetic Field Laboratory (NHMFL), Los Alamos and the 60 T pulsed field magnet in the National High Magnetic Field Center, Wuhan.

The change was measured in the resonant frequency of a dissipationless LC-circuit formed by a measurement coil with the sample attached to it. The change in resistivity of the sample changes the inductance of the measurement coil, which in turn shifts the resonant frequency of the circuit. A measurement coil with 5-6 turns was employed, and PDO circuits with a typical resonant frequency of 30 MHz were chosen for optimal signal-to-noise conditions. The resonant frequency was measured after a signal processing stage with a National Instruments digital oscilloscope.

## Data availability

The data represented in Figs. 2-4 are available on the University of Cambridge data repository (https://doi.org/10.17863/CAM.50169). All other data that support the plots within this paper and other findings of this study are available from the corresponding author upon reasonable request.

## Acknowledgments

M.H., Y.-T.H. and S.E.S. acknowledge support from the Royal Society, the Winton Programme for the Physics of Sustainability, EPSRC (studentship, grant number EP/P024947/1, and EPSRC Strategic Equipment Grant EP/M000524/1) and the European Research Council (ERC Grant Agreement number 772891). S.E.S. acknowledges support from the Leverhulme Trust by way of the award of a Philip Leverhulme Prize. H.Z., J.W. and Z.Z. acknowledge support from the National Key Research and Development Program of China (Grant no. 2016YFA0401704). A portion of this work was performed at the National High Magnetic Field Laboratory, which is supported by the National Science Foundation Cooperative Agreement No. DMR-1644779, the state of Florida, and the U.S. Department of Energy. Work performed by M.K.C., R.D.M. and N.H. was supported by the US DOE BES 'Science of 100 T' program.

## Author contributions

M.H., Y.-T.H., K.A.M., H.Z., J.W., Z.Z., M.K.C., R.D.M., S.E.S. and N.H. performed high-

magnetic-field measurements. Y.-T.H., J.P., T.L., M.L.T. and B.K. prepared single crystals. M.H., Y.-T.H., R.D.M., G.G.L., S.E.S. and N.H. contributed to data analysis. S.E.S. and N.H. conceived the project. S.E.S. and N.H. wrote the manuscript with M.H. and Y.-T.H., with contributions from all the authors.

## Competing Interests

The authors declare that they have no competing financial interests.